# How Far Can We Go Through Social System?
Algorithmic Information Social Theories

Hokky Situngkir
(hokky@elka.ee.itb.ac.id)
Dept. Computational Sociology
Bandung Fe Institute

**Abstract**
The paper elaborates an endeavor on applying the algorithmic information-theoretic computational complexity to meta-social-sciences. It is motivated by the effort on seeking the impact of the well-known incompleteness theorem to the scientific methodology approaching social phenomena. The paper uses the binary string as the model of social phenomena to gain understanding on some problems faced in the philosophy of social sciences or some traps in sociological theories. The paper ends on showing the great opportunity in recent social researches and some boundaries that limit them.

## 0. Prologue

Meta-sociology is the term to the inquiries of philosophical issues circling social sciences, the way to build the sociological theory, and the evolution of social sciences regarding the methodology and fundamental principles motivating them. As 1931 saw the issue of incompleteness theorems sounded by Kurt Gödel, they have influenced many parts of scientific domains; however, the social sciences seemed to have been ignoring the important impact. It happened probably because the incompleteness theorems are elaborated in rigorous language to the rigorous arithmetic system. In this perspective, the paper elaborates an endeavor on applying the algorithmic information-theoretic computational complexity – as an effort to see impacts of Gödelian incompleteness theorems to the reality in broader view – to meta-social-sciences.

The paper tries to be firmed and not that rigorous. A glance reader can read the paper in the terms of "#" and "*" denoting "definition or axioms" and "deductively causal impact" established in the paper respectively. The paper is structured by noting some definitions considering the construction of social theories, followed by the result or consequence of the implementation of algorithmic information theory to explain some traps that sociological theories face. Aftermath, the paper continues on elaborating recent issue on some contemporary methodology in social sciences.

**1. The Construction of Social Theories**

Social sciences are in some epistemological aspects standing for the similar scientific methodology commonly known in the natural sciences. The idea is to grasp the patterns we find in the social phenomena in order to have a sound explanation on them. Qualitatively, we saw a phenomenon as empirical inquiries and then have it analyzed to construct the social theory. The social theory can be regarded as quasi-axioms on understanding many social phenomena. In fact, this procedure could also be thought as finding the reducibility of the social phenomena seemingly random we found in the real life.

The social theory is by means of the **explanation** on social phenomena. As stated by classical sociologist, Emile Durkheim (1895:97), the results of the preceding method concluded as social theory should need verification by demonstrating that the general character of the phenomenon is related to the general conditions of collective life in the social type under consideration. The social theory is meant to be the information that enable us to generate the social phenomena in sociological thoughts in order to have understanding.

**# 1**
*Social phenomena are the observable certain events within social system.*

In this case, the term 'social phenomena' includes all the events involving identity (individual and collective), culture, symbols, ideas, norms, principles, narratives, and collectively held beliefs. The events can be certain occasions, for example social conflict, voting process, rituals, consumptions, etc, while the set of social phenomena can be regarded as an infinite set.

The way to construct social theory is to *reduce* the information described as social phenomena. The well-known dictum of scientific endeavor, Occam's Razor fits this situation. Occam's Razor principles stated that: given some theories of equal merit, the simplest is to be preferred. As commonly sociologist, Wheaton (2003) felt Occam's Razor as a form of oppression. There is a dilemma whether the theory with better parsimony must have the better explanation. In this case, we can moderately concur that **the one that applies most generally must have the best explanation**. The use of Occam's Razor or the principle of reducibility in social science must be established in this perspective. The simplest social theory is the one that applies the most generally.

The idea that there is a "general social theory" enabling us to explain all social phenomena was felt much among some social theorists. A remarkable sociologist, Talcott Parsons (1951) contended that to be a sociologist is to be a theorist pursuing the abstract general theory on explaining as many as possible occasions of social phenomena. The spirit of Parsons sounds similar to the Hilbert's tenth problem of the task of mathematics to construct the absolutely consistent formal axiomatic system in its entirety – every deductive reasoning can be processed by the formal axiomatic system (Nagel & Newman, 2001:27). However, Kurt Gödel showed the incompleteness theorem implying that for any given finite theory of the universe, there are certain facts having to do with sets of physical objects that cannot be proved by the theory (Sullins, 1997).

However, most of social theorists to day are against Parsons on this claim. An alternative was presented by Robert K. Merton (1968:39) by the notion of the theories of middle-range – some theories constructed by guiding the empirical researches synergistically with expectation on explaining all kinds of social phenomena. In this case, Merton offered a way to bridge the general theory and special observed occasions: a method from the deduction with the other needed empirical inquiries. The general theory of social system can only be gained by certain generalizations among empirical findings.

In this sense, we understand that the way of establishing social theory is the way to compress the information gained in social phenomena, as parsimonious as possible, by means of explaining the most generally.

# # 2

*Social theory is the set of compressed social phenomena whose highest parsimony (relative to their most applicability). The social theory explaining the social phenomena is recognized as the way to reconstruct or decompressed the certain phenomena from certain social theory.*

Just as concurred by Durkheim(1895:134-135), the determining cause of a social fact must be sought among the antecedent social facts and not among the states of the individual consciousness; that the function of a social fact must always be sought in the relationship which it bears to some social end.

Ray Solomonoff (1975) presents a famous model for the scientist's observations in a series of binary digits. He shows that the endeavor of the scientist to explain her observations through a theory can be regarded as an algorithm capable of generating the series and extending it, that is, predicting future observations. However, there would always be several competing theories, forcing the scientist to choose among them. According to Occam's Razor, the model demands that the smallest algorithm, the one consisting of the fewest bits, be selected in the principle of that **the more comprehensive theory should have greater degree of explanation** (Chaitin, 1975).

The question emerging from Solomonoff's model and the compressibility of Chaitin is how far we can compress the social phenomena. As explained clearly by Chaitin (2004), we can apparently compress certain bits of binary digits, however, there are some bits that cannot easily be compressed by the notion of the randomness in Algorithmic Information Theory. Information embodied in a random series of numbers cannot be ``compressed,'' or reduced to a more compact form.

Continuing Solomonoff's model, any minimal program is necessarily random, whether or not the series it generates is random. Consider the program *T*, which is a minimal program for the series of digits *P*. If we assume that *T* is not random, then by definition there must be another program, *T'*, substantially smaller than *T* that will generate it. Apparently, we can produce *P* by the following algorithm: ``From *T'* calculate *T*, then from *T* calculate *P*.'' This program is only a few bits longer than *T'*, and thus it must be substantially shorter than *T*. *T* is therefore not a minimal program (Chaitin, 1975).

As understood to be Laplacean isomorphism, we recognize that computation does not provide a map of the universe; however, the universe is a map of a computation (Patee, 1995). This proposition simply permits us to model a social theorist as a particular theoretical computer, keep seeking the explanation of social phenomena based on his understanding on social theories[1]. However, the theoretical computer should be realized as a partial recursive function since there is no guarantee that the computer will always have ability to explain certain phenomena[2].

---

[1] We do not concern about how to model cognitive system as a computation here since it will bring us to the debate initiated by some hypotheses of Penrose (1994:12-7) Gödel's Incompleteness Theorem. What we are trying to explain is when the theorist use the logical rule of inference on her explanation about the social phenomena based on theories. This claim came from the principles of scientific method on logical

<div style="text-align:center">

social research ≡ social theories → theorist → social phenomena

program → computer → output

</div>

**# 3**
*The way to do social research is to explain such social phenomena by doing certain algorithm by means of rules of inference on social theories to reconstruct the respective social phenomena. Theorist and a theoretical computer is assumed isomorphism, in the condition of the respective theoretical computer is partial recursive.*

By adopting the formal definition of Chaitin (1974) and the above definition, we can assume that the way to explain a particular phenomenon from certain theory as a partial recursive process $T: S \times S \to S$, $T(a,h)$, where $a$ is a binary string and $h$ is a particular natural number representing length of the explanation, with property that $F(a,h) \subseteq F(a,h+1)$. The value of $F(a,h)$ is the finite set of phenomena that can be derived from $a$ by means explanation $\leq h$ characters length. In this case, $F(a) = \bigcup_h F(a,h)$ is the set of phenomena as logical consequences of theories $a$. As in **#3**, we defined that the phenomenon analyzed should be of best parsimony, hence the complexity measure of a phenomena represented as binary strings $s$ should be denoted as $H(s) = \min_{T(p)=s} length(p)$, and if $T(p) \approx \varnothing$ then $H_T(s) \approx \infty$; recognized as the maximally complex social phenomenon. Following Chaitin, we can imagine universal computation procedures to measure the complexity in which $H_U(s) \leq H_T(s) + c$ where $c$ depends on the language used in describing the theory.

## 2. The logical incoherence as a trap in Sociology

From the definitions above, we can see the explanation of social phenomena based on particular social theories as a computational process. In this section, we will see some corollaries of Chaitin's algorithmic information theory to the social analysis.

**\* 4  Chaitin's Theorem**
*There exists constant c that fulfils the value of complexity of certain s bit strings less than addition of the length of s and c*

The proof of the above theorem can be seen in Chaitin (1974), giving us the thesis that since $H(s) \leq length(s) + c$, then for most of big size binary strings *s*, $H(s) \approx length(s)$. By using our analogy in the previous section, we have,

---

coherence and some flaws frequently existing in social theories. We left this to the next section of the paper.

[2] A function is recursive if there is an algorithm for calculating its value when one is given the certain value of its arguments, in other words, if there is a Turing machine for doing this. If it is possible that this algorithm never terminates and the function is thus undefined for some values of its arguments, then the function is called partial recursive (see Martin, 1991:370-377).

**\* 5** (Thesis)
*In most of social researches, the theoretical complexity of phenomena will tend to be similar to its size.*

This is a direct corollary we have from the model of social researches – modeled as in the previous theorem (**\*4**). If we have *s* bit strings as an oversimplification of social phenomena, then we could find out that the complexity description will always be $H(s) \leq length(s) + c$, where *c* depends on the language use to have it. In fact, this has been recognized by Craib (1992:10-13) as one of the prominent traps found in sociological theory, the 'description trap',

> "This can be understood by distinguishing the term 'explanation' and 'description'. While explanation means telling something cannot recognize simply by looking, e.g.: patterns, even prediction; and in the other hand description tells things discovered by looking. However, it seems to be often that thousands of theoretical pages in social analysis to be translated on nothing but jargons. The works on postmodern sociology is vulnerable on this point, while there are thousands of best-selling and well-known books rush the market but add little or even nothing to our knowledge."

This also has been qualitatively showed by Situngkir (2003) that the social theories should then be applied in respect to the particular social space, thus the social theories would be very sensitive to the system they approach. Most of social theories are then expected to be able to be spatio-temporal and there is no scientific reason of using the theories taken from assumptions in different localities to be general. Eventually, this also brings us a denial to the propositions expected by the Parsonian sociology as described in the previous section.

**\* 6 Lemma from Algorithmic Information Theory**
*It is impossible to prove that a certain binary string is of complexity greater than n+c in a formal system with n bits of axiom. However, there are formal systems with n+c bits of axioms that possible to determine whether the complexity of a string is less or greater than n, but without ability to prove how much the complexity of the string exceeds n.*

Consider we have a proposition of the '$H(s) \geq n$' in a rules of inference *F(a)*, Chaitin's theorem contended that such proposition is in *F* only if $n \leq length(a) + c$ where *c* is a constant depends only on *F*. The proof of the theorem can be accessed in Chaitin (1974). Consideration to this brings us to major facts that although the social theory is intended to find logical coherences among theories and empirical facts, there exists some of the theories are illogical. However, some theorists proposed that it came from the fact that some sociological phenomena are not at all coherent logically.

**# 7**
*X set of elements S is said to be infinite if the elements of a proper subset S' can be put into one-to-one correspondence with the elements of S. In this case, two sets can be put into one-to-one correspondence if and only if their members can be paired off such that each member of the first set has exactly one counterpart in the second set, and each member of the second set has exactly one counterpart in the first set.*

In the language of algorithmic information theory, the infinite complexity is by meaning of maximally unknowable, which will never be fulfilled in a compression of information. The whole will always be much bigger than the part. Thus, the length of information cannot be measured precisely. In advance, some fruitful discussions of infinity can be seen in Suber (1998).

# 8
*The complexity of the social phenomena is infinite.*

* 9 (Thesis)
*The complexity of sociological thought is infinite.*

The definition of **#7** can simply be thought as a tautology with our previous axioms on modeling the sociological thought as partial recursive system, where there is a probability that the theoretical computation does not halt. However, by the lemma **\*6** we can simply find that there will always be possibility to decide whether or not a binary string exceeds or not the length of the theoretical computation. This is emphasized by our thesis **\*5**. Furthermore, by the definition **#7**, we can also simply demonstrate that the complexity of social theory can always be put into one-to-one correspondence with the elements of itself. Even though the social theory cannot be seen as compressed social phenomena to be explained, the resulting description could be recognized as the theory occasionally.

* 10 (Corollary)
*It is impossible to capture social phenomena completely into a single social theory in the very logical circumstances, i.e.: consistency and completeness as a whole.*

Eventually, we can have now the widened Gödelian incompleteness theorems to the social sciences by using the algorithmic information theory that the social theory will never be able to capture the social phenomena as a whole into the social theory based upon deduction by means of explaining them based on the existing theories in the fashion of formal axiomatic system. In other words, there should be a synergetic approach among the logical coherence in deductive inference with the inductive one as elaborated in the next section.

* 11 (conjecture)
*The power of social theories can be measured on two fashions, i.e.: the power of explainability and the predictability.*

The above conjecture simply can be regarded as the following propositions came from the major question on meta-social-theory. As noted by Merton (1945), the paradigm of "proof through prediction" is categorized to be fallacious in social theorem. This is another version of the challenge of theory constructions in social sciences beside Coleman's (1990) emphasizing the importance of explanation more than prediction in social sciences. There are certain theories of comprehension purporting prediction and in the other hand; there are necessity on explanation more than just a prediction. Nonetheless, we should note that practically social sciences have more than just the explanation and prediction. Collins (1998) noted that,

> …sociology is nearly the most politicized and activist of all fields….Probably the
> only disciplines that are even more thoroughly politicized than sociology are

> relatives of sociology, e.g.: ethnic studies, black studies, and women's studies, which were created as hybrids between academic departments and activist movements.

This becomes bias in the practical use of social theory that came from the nature of the objects being analyzed. While in this case, the entirety information of social theories remains nothing but fuel in political and ideological debates, and certainly becomes out of our discussion about the meta-analysis of social theory.

**3. Philosophical and Practice of Social Research**

The way to do social research is in short can be described as the way to do the deductive reasoning and empirical inquiries synergistically. However, conventional research method relied merely upon statistical analysis cannot cope with our enlargement of the complexity of social research respect to our previous elaboration. The recent computational techniques are introducing the way to have it in the informally branch of sociology, i.e.: computational sociology. The using of computational sociology is the matter of inter-playing the social factors with the social actors (Macy & Willer, 2002). As pointed out by Gilbert & Troitzsch (1998), the method can be assessed by following the scheme in figure 1.

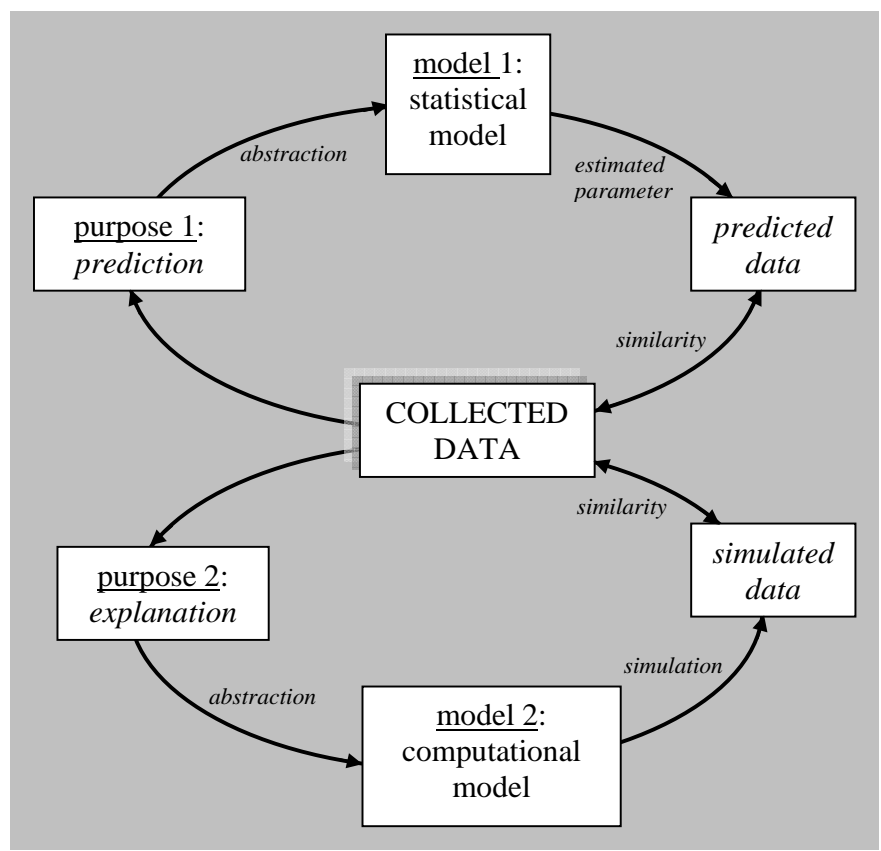

**Figure 1**
The different purposes on prediction and explanation in social research.

It is obvious that the abstraction to constructing the model can distinct the model we are dealing with. Furthermore, the computational model we use to explain social phenomena can factually discovered in the recent methodology of artificial society. In the analysis with artificial society, we capture the social process with its statistical properties and construct the agent-based model emerging the best proximity with the former. There have been many practical discussions in this issue, e.g.: Doran (1997), Axtell (2000), Langton ((2002), as the term coined by Epstein & Axtell (1996).

Eventually, we have figured out that there is wide-open opportunity on capturing the social phenomena with compact and explainable theory by taking advantage from the computational technology while in return we realize how the boundaries we have on every possible social research and theory construction through advanced understanding on computation. These are two quite different uses of the computation.

## 5. Concluding Remarks

We present the way to widen the impact of incompleteness theorem by using the algorithmic information theory as originally introduced by Chaitin (1974). We present a model of meta-social-research to be partially recursive computation and show some important impacts on opportunity and boundaries in endeavor of social sciences. In parallel, we show that it is very difficult capturing the social phenomena and explain them using the existing social theories. This is, however, the nature of social sciences as compared to natural sciences.

There is also a tendency to be descriptive over social sciences by realizing the uniqueness of social phenomena (highly random social phenomena). However, this tendency – e.g.: postmodern sociology – can be damped by our practical computational technology enabling us to sharpen the way as to construct social theories for a theory just as like a report without ability to explain (or in reverse as a compression of phenomena) remains useless.

Furthermore, by applying the algorithmic information theory into social sciences we have newly understanding about the occasional traps of sociological theories, i.e.: the description trap and logical trap. We show that most of social phenomena are maximally unknowable with infinite complexity of social phenomena contrasted to the infinite complexity of social theory. As depicted before, the infinite complexity of social phenomena leaves us boundaries but in return, the infinite complexity of social theories brings us a great deal of opportunity on denouncing the unknowable rest for us for making a better social living.

**Acknowledgement**

The author thanks the Surya Research Intl. for financial support in which period the paper is written, Yohanes Surya for some important literatures, and the academic society of BFI in which the paper have been reviewed.